\documentclass[letterpaper]{article} 
\usepackage{aaai2026}  
\usepackage{times}  
\usepackage{helvet}  
\usepackage{courier}  
\usepackage[hyphens]{url}  
\usepackage{graphicx} 
\usepackage{natbib}  
\usepackage{caption} 
\usepackage{array}
\usepackage{booktabs} 
\usepackage{tabularx}
\usepackage{algorithm}
\usepackage{algorithmic}

\usepackage{newfloat}
\usepackage{listings}
\DeclareCaptionStyle{ruled}{labelfont=normalfont,labelsep=colon,strut=off} 
\floatstyle{ruled}
\newfloat{listing}{tb}{lst}{}
\floatname{listing}{Listing}
\title{AI From the Margins (AIM): Rethinking Participatory AI Design Through the Lived Experience of Minoritized Communities}
\author{
    Tijs Portegies\textsuperscript{\rm 1, 2}, Laureanne Willems\textsuperscript{\rm 3}, Maaike Harbers\textsuperscript{\rm 2}, Giovanni Sileno\textsuperscript{\rm 1}, Roland van Dierendonck\textsuperscript{\rm 2}, Mayesha Tasnim\textsuperscript{\rm 1}, Lotte Willemsen\textsuperscript{\rm 1, 2}, Sennay Ghebreab\textsuperscript{\rm 1}\\
}
\affiliations{
    \textsuperscript{\rm 1}University of Amsterdam\\
    \textsuperscript{\rm 2}Rotterdam University of Applied Sciences\\
    \textsuperscript{\rm 3}University of Toronto\\

}

\usepackage{bibentry}

\begin{document}

\maketitle

\begin{abstract}
Artificial intelligence (AI) can reproduce and amplify the structural inequities faced by minoritized communities. Participatory AI has been proposed as a response, but participation typically starts after problem definitions and success criteria have been set, leaving limited room for minoritized communities to reshape what an AI system is for. We propose AI From the Margins (AIM): a methodological stance that articulates the conditions under which lived experiences of minoritized communities can be elicited, centered, and carried forward to inform participatory AI design. AIM is not a fixed protocol; it articulates a set of preconditions that can be enacted through different techniques in different settings. We applied AIM in a Dutch healthcare context in eight sessions with 13 women and non-binary people of color and five municipal policy workers, namely through (1) narrative elicitation using the Biographic Narrative Interpretive Method (BNIM); (2) co-constructed rule-making; (3) participants' determination of whether, where, and how AI should be involved; and (4) translating lived experience into AI policy through dialogue with policymakers. In their reflections on the sessions, participants described the engagement as substantive and called for its continuation, demonstrating how preparatory orientation fundamentally grounded in lived experience shapes what participatory AI design is for.
\end{abstract}


\section{Introduction}
Artificial Intelligence (AI) is increasingly embedded in several domains of human activities. In healthcare, for instance, it shapes clinical decision-making, disease detection, support for daily living, and biomedical data analysis \cite{Rong2020}. Despite its promise to optimize healthcare, research shows that AI applications can reproduce and even amplify existing inequities, including racial bias and discrimination, creating particular vulnerabilities for \textit{minoritized} patient populations \cite{Cho2024}.\footnote{Minoritized communities are defined as groups actively positioned at the margins through systemic processes, not due to inherent minority status but as a result of structural exclusion within dominant healthcare institutions and systems \cite{Quyoum2024}.} Prior scholarship has shown that algorithmic systems can mirror deeply ingrained racial biases, including those historically embedded in medical beliefs about pain and treatment in racialized patients \cite{Akinlade2020}. For example, a recent study found that, when asked for medical advice, Large Language Models (LLMs) discriminate against users who are Black, Queer, unhoused, or poor \cite{omar2025}. 

Participatory AI has emerged as a prominent response to such harms, promising to involve affected communities in shaping the systems that affect them \cite{Sloane2022}, in practice aiming to realize principles of \textit{discourse ethics} \cite{Mingers2010}. However, much of what is labeled participatory remains performative or instrumental, with participation invoked to legitimize systems that have already been designed rather than to share authority over the design process \cite{Delgado2023}. In light of this critique, studies emphasize that participatory AI must go ``beyond general engagement and center the lived experiences of the most affected'' \cite{Birhane2022}. These claims are grounded in \textit{Design from the Margins} \cite{CostanzaChock2020}, which advocates starting design processes with those most marginalized by existing systems, whose experiences reveal structural conditions that require transformation. In her work, \citet{Crenshaw1991} explains this effect with her concept of \textit{intersectionality}: race, class, gender, and other systems of power shape lived experience in ways that cannot be understood along a single axis, so design frameworks must engage minoritized identities holistically rather than as isolated variables. \citet{Quyoum2024} translate this into AI specifically, calling for a shift from treating minoritized individuals as users to engaging them as active participants in shaping technological systems. Together, these positions establish why the lived experiences of minoritized communities are not simply one input among many, but the standpoint from which AI's harms become most visible \cite{Ajmani2024, Birhane2022, collins2000black}. Yet, concrete guidance on systematically incorporating lived experience into AI development remains limited.

Recent work in participatory AI argues that incorporating lived experience into AI development requires more than expanding who participates: it requires methodologies capable of engaging situated and experiential knowledge within the social contexts in which AI is encountered \cite{Katell2020, Suresh2024}. In most participatory AI processes, however, participation is bounded before it begins. \citet{Delgado2023} show that stakeholders typically enter a process whose problem definitions, success criteria, and the design of the AI applications are already set, leaving little room for their contributions to reshape what the system is for. \citet{corbett2023power} sharpen this point: control over problem framing is itself a site of power, and participation invited into an existing frame cannot redistribute the power exercised in setting it. Standpoint theory \cite{collins2000black} clarifies why this matters epistemically. Knowledge from the margins surfaces the operations of power and exclusion that remain illegible from the center, which means that only elicitation beginning outside the dominant technical frame can unsettle it. Taken together, these arguments point to a methodological need that existing participatory AI currently does not meet: a stage that both precedes technical framing and foregrounds the intersectional lived experience of those most affected by AI applications' harms. \citet{portegies2026bnim} begin to address this by applying (as exploratory device to envision sound participatory AI requirements) the Biographic Narrative Interpretive Method (BNIM), a qualitative method in which participants structure their accounts around events they themselves consider significant, reducing researcher-imposed framing and surfacing dimensions of lived experience that structured elicitation tends to leave underarticulated. Their findings show that a pre-design stage grounded in BNIM can generate both recognized and novel requirements for AI in healthcare \cite{portegies2026bnim}. Yet BNIM has not been articulated as a systematic preparatory stage for participatory AI, nor have its outputs been integrated into a method explicitly meant to precede and shape participatory AI design.\footnote{Preparatory here means prior to any technical framing: AIM is completed before the AI system's purpose, scope, or design criteria are defined.}

To address this gap, we propose ``AI From the Margins'' (AIM): a methodological stance that specifies a preparatory stage prior to participatory AI design, grounded in standpoint theory, decolonizing research, and the Design from the Margins tradition. Rather than prescribing a fixed protocol, AIM articulates a set of preconditions, commitments that must be in place for participatory AI to genuinely center the lived experiences of minoritized communities. These preconditions are not procedural steps to follow; they are standing requirements, without which participation risks reproducing the harms it sets out to address. Together, the seven preconditions we propose position AIM not as an additional participatory AI pipeline to be added to existing methods, but as the formulation of what any such methods must do to center and take the lived experiences of minoritized communities seriously. 

Our research question is: \textit{How can lived experience be operationalized as a foundation for participatory AI design with minoritized communities in healthcare?} This overarching question gives rise to several sub-questions: on context (preconditions which are necessary for proper participatory AI design to occur), on meta-process (how to initially set up the process, or with which rules participants can modify the process itself), on process (what happens when a specific instance is executed), on outcomes. The hypothesis motivating the introduction of AIM is that preconditions provide the necessary backdrop against which genuine participation can unfold, and for this reason are the first frontier to address. Because of its contextual nature, AIM can therefore be seen as a meta-method. At a conceptual level, AIM has been constructed by systematizing and elaborating on relevant literature and previous work. Its practical validation requires in principle a two-step process: first, applying AIM to identify a context-dependent method for participatory AI, and second, executing the established method. We therefore applied AIM in the Dutch healthcare domain, conducting eight Lived Experience Sessions, each designed to enact AIM's preconditions through specific techniques (SQUIN, Rich Picture, dialogue with policy workers), with 13 women and non-binary people of color.

At this point, effects can be assessed based on the participants' perspectives on the overall process (\textit{procedural success}) and based on outcomes (\textit{downstream success}). Yet, the present work concerns primarily the development and articulation of AIM as a methodological stance; 
the empirical application reported here serves as a worked example through which AIM's preconditions are enacted and illustrated, rather than as a study designed to test or evaluate the method's general effectiveness.

The paper proceeds as follows. We first review related work on participatory AI and lived experience, then introduce the AIM preconditions and describe their execution in the Lived Experience Sessions, before presenting empirical findings and theoretical implications.

\section{Related Literature}

\subsubsection{Participatory AI and Minoritized Communities}

``[I]f AI is to protect the welfare and well-being of the most negatively impacted stakeholders, it needs to be guided by the lived experiences and perspectives of such stakeholders'' \cite{Birhane2022}. 
In the context of AI, lived experiences of minoritized communities provide crucial insights into how AI applications fail, where harms are perpetuated, and what kinds of justice and accountability are needed. Building on this, \citet{Sloane2022} propose the concept of \textit{participation as justice}, which centers long-term partnerships with diverse stakeholders grounded in mutual benefit, reciprocity, and equity. Their argument draws on previous work by \citet{hooks1994}, who analyzes reciprocity as a deliberate break from relationships structured by hierarchy and domination, especially between people in different positions within systems of power. Translated into participatory AI, this means that participants must not be treated as ``subjects to be studied'' or ``sources of data'', but as \textbf{co-owners} in the production of knowledge \cite{hooks1994, Sloane2022, Sum2025}. This aligns with \citet{TuckYang2014}'s articulation of \textbf{refusal} as a legitimate methodological outcome. In this account, \textbf{reciprocity} is not a procedural courtesy but a relational orientation that participatory AI must build rather than assume.

If reciprocity describes the relation that participatory AI must establish, power can be described as what stands in the way of reciprocity. \citet{corbett2023power} argue that meaningful participation in AI design must address the broader social structures that condition who controls the process, who makes decisions, and how participants' knowledge is used. \citet{hooks1994} argues that non-oppressive relationships depend on honest confrontation: not the polite acknowledgment of inequality, but the direct surfacing of unequal power between people in different positions within systems of domination. Translated into participatory AI design, honest confrontation requires asking, and answering, uncomfortable questions about the participatory framework itself: who holds power within the process, whose decisions shape its terms, and whether participation is being used to transform a system or merely to optimize and legitimize it \cite{Delgado2023}. \citet{Udoewa2022} sharpens this critique in his account of radical participatory design, arguing that participatory practices can themselves reproduce colonial extraction when institutions retain control over framing, success criteria, and the terms on which participant knowledge is made legible. What participatory AI requires, then, is not only better methods, but a deliberate redistribution of who decides what participation is and a willingness to forgo predetermined framings in favor of participant-led direction, emphasizing \textbf{methodological flexibility} \cite{Udoewa2022, Kallina2025}. \citet{Sloane2022} describe what this looks like in practice, noting that genuine depth arises when participants and facilitators can identify and explore (uncomfortable) dynamics, including power, privilege, and institutional patterns, rather than smoothing them over for comfort or cohesion. 

Reciprocity and the redistribution of power both depend on a deeper epistemic claim: the lived experiences of minoritized communities are not one input among many, but a particular standpoint from which the workings of AI applications become legible \cite{Birhane2022, Sloane2022}. \textit{Standpoint theory} \cite{collins2000black, harding1991} provides the warrant for this claim. More specifically, Collins (2000) introduces the concept of the outsider-within, which names the position from which those structurally excluded from dominant institutions can both see those institutions clearly and recognize the operations of power that remain invisible to those at the center. Marginalized perspectives should not be treated as homogeneous; \citet{Crenshaw1991}'s concept of intersectionality demonstrates that lived experience is shaped by multiple, intersecting systems of power, so that the standpoint of, for example, a Black queer patient navigating healthcare cannot be reduced to that of either Black patients or queer patients in the aggregate. \citet{CostanzaChock2020}'s Design from the Margins translates this convergent epistemic claim into a justice-oriented design framework, arguing that design processes must begin with those most marginalized by existing systems because their lived experiences expose the structural conditions that any meaningful design must address. For AI design, this convergence carries a methodological implication.  
What is required is not better proxies for marginalized identity in datasets, but methodologies capable of \textbf{centering minoritized standpoints} \cite{Eslami2025}.

\begin{table*}[t]
\centering
\caption{The seven preconditions of AIM: definitions and theoretical grounding.}
\label{tab:aim-preconditions}
\small
\setlength{\tabcolsep}{5pt}
\renewcommand{\arraystretch}{1.3}
\newcolumntype{L}{>{\hsize=1.4\hsize\raggedright\arraybackslash}X}
\newcolumntype{R}{>{\hsize=0.6\hsize\raggedright\arraybackslash}X}
\begin{tabularx}{\textwidth}{>{\raggedright\arraybackslash}p{3.2cm} L R}
\toprule
\textbf{Precondition} & \textbf{Definition} & \textbf{Theoretical Grounding} \\
\midrule
Reciprocity
& A relational commitment in which participants and researchers are co-producers of knowledge, mutually accountable for what is generated.
& \cite{hooks1994}; \cite{Sloane2022}; \cite{Sum2025}. \\
\midrule
Decolonizing stance
& Participants are positioned as agents of inquiry with authority over which questions are worth asking, what counts as relevant knowledge, and how findings are shared and used.
& \cite{smith2012}; \cite{dignazio2020data}. \\
\midrule
Centering minoritized standpoints
& The lived experiences of minoritized communities are treated as an epistemic standpoint from which harms (of AI) become legible, rather than as one input among many.
& \cite{collins2000black}; \cite{harding1991}; \cite{Crenshaw1991}; \cite{Eslami2025}. \\
\midrule
Methodological flexibility
& Forgoing predetermined framings in favor of participant-led direction, while maintaining methodological coherence.
& \cite{Udoewa2022}; \cite{Kallina2025}. \\
\midrule
Structural availability of refusal
& Participants retain the right to refuse aspects of the research, including AI itself as an appropriate response, treated as a legitimate methodological outcome rather than a failure.
& \cite{TuckYang2014}; \cite{smith2012}. \\
\midrule
Multidimensional accessibility
& Material conditions for participation including timing, location, food and care, and accommodations responsive to participants' specific needs.
& \cite{CostanzaChock2020}; \cite{Kuo2023}. \\
\midrule
Co-ownership
& Participants’ control over the research process and its outputs, including transparency regarding purposes, access to data, withdrawal, and findings.
& \cite{Sloane2022}; \cite{smith2012}; \cite{Tseng2025}. \\
\bottomrule
\end{tabularx}
\end{table*}

\subsubsection{Methodological Frameworks for Participatory AI and Lived Experiences}
The political and epistemic terms of research with minoritized communities cannot be separated from the methods themselves \cite{dignazio2020data, smith2012}. Smith (2012) argues that research has historically functioned as an instrument of colonization, producing knowledge about Indigenous and marginalized peoples that has been used against rather than for them, and that taking a \textbf{decolonizing stance} requires minoritized communities to become agents of inquiry with authority over what questions are asked, what counts as knowledge, and how findings are used. This also takes into account the material conditions for participation, including timing, location, food and care, and accommodations responsive to participants’ specific needs \cite{CostanzaChock2020, Kuo2023}. In her work, Collins (2013) develops \textit{intellectual activism} as a methodological stance that holds scholarly work accountable not only to academic institutions, but to the communities it concerns, recognizing that methodological choices are political choices that determine whose frameworks prevail and who benefits. The methodological challenges of participatory AI design can therefore be understood not merely as technical questions, but as questions of \textit{accountability} and \textbf{epistemic authority}. Such concerns should orient any framework that claims to center the lived experiences of minoritized communities \cite{collins2000black}. 

Recent work in AI design has begun to address this challenge. Most notably, the \textit{Lived Experience AI Framework} (LEAF), presented by \cite{gautam2025experience}, synthesizes interdisciplinary literature on lived experience and offers a taxonomy with stages designed to be integrated into the AI design and development life cycle. LEAF makes an important contribution in establishing lived experience as a legitimate design concern and in mapping its dimensions across multiple application domains, including healthcare \cite{gautam2025experience}. However, several features of LEAF reflect a different orientation from the one developed in this research. First, LEAF centers pluralistic lived experience as broadly enriching to design, while the present work, drawing on standpoint theory and intersectionality, treats the lived experience of minoritized communities as a particular epistemic standpoint rather than one perspective among many. Second, LEAF integrates lived experience within the AI design pipeline, where, as \citet{Delgado2023} and \citet{corbett2023power} explain, problem framings, success criteria, and design artifacts are typically already set before participants enter the process. AIM, in contrast, is designed to precede these framings, so that lived experience can shape \textit{what the design is for} rather than how an existing design is refined. Finally, while LEAF references existing participatory methods at each stage of the AI development pipeline (e.g., co-design workshops, community-led data annotation), AIM offers a methodological stance whose preconditions can be enacted through various techniques across different settings. In this paper, in particular, we demonstrate AIM through structured biographic narrative sessions grounded in BNIM \cite{Wengraf2001, portegies2026bnim}, designed for the preparatory stage.

\section{Method}
At a fundamental level, AI From the Margins (AIM) is meant to operationalize lived experience as the center of participatory AI, progressing from narrative to design and governance. AIM systematizes into a cohesive whole seven preconditions presented in diverse contributions (Table~\ref{tab:aim-preconditions}): (i) \textit{reciprocity} as a relational commitment, rather than a procedural one  \cite{hooks1994, Sloane2022}; (ii) a \textit{decolonizing} stance in which participants are agents of inquiry rather than objects of study \cite{smith2012}; (iii) the \textit{centering of minoritized standpoint} as foundational input rather than one input among many \cite{collins2000black, Crenshaw1991, harding1991}; (iv) \textit{methodological flexibility}, including a willingness to forgo predetermined framings in favor of participant-led direction \cite{Udoewa2022}; (v) \textit{structural availability of refusal}, including the refusal of AI itself as an appropriate response \cite{TuckYang2014}; (vi) \textit{multidimensional accessibility}, which attends to the material conditions under which participation becomes possible \cite{CostanzaChock2020}; and (vii) \textit{co-ownership}, in which participants retain control over the research process and the data it produces \cite{Sloane2022, Birhane2022, smith2012}. 

Several approaches can be applied to meet these preconditions, with their effectiveness most plausibly depending on the context. It is not our objective here to find the best ones, but rather to demonstrate the possibility of AIM's realization. We therefore describe below one application of AIM's preconditions, unfolded across eight sessions. Note that while the specific techniques are substitutable, the preconditions themselves are not.

\subsection{Design of the AIM sessions}

\subsubsection{Session 1: Sharing Lived Experience in Healthcare}

We designed Session 1 to position lived experience as AIM’s starting point, foregrounding reciprocity and co-ownership to establish the relational foundation for subsequent sessions. The structure of the session was guided by the Biographic Narrative Interpretive Method (BNIM) \cite{Wengraf2001}, which begins with a Single Question aimed at Inducing Narrative (SQUIN) to elicit participants’ lived experiences. In this case, we asked each participant: \textit{``Please tell me in your own words about all the events and experiences with healthcare that have been personally important to you?''} BNIM has been shown to be useful in the context of lived experience research \cite{portegies2026bnim}. To deepen the discussion, we incorporated a Socratic dialogue–inspired approach to encourage participants to ask one another follow-up and probing questions, helping them gain a fuller understanding of each other’s lived experiences \cite{harbers2019socratic}. 

The opening procedures of Session 1 enacted reciprocity and co-ownership as relational commitments rather than procedural steps. Drawing on hooks (1994), Sloane et al. (2022), and Birhane et al. (2022), the session began with an explicit discussion of participants’ rights to privacy and to communicate boundaries, their right to access session recordings, and a transparent explanation of who else would have access. The session leader addressed any questions or concerns before starting the recording and asked participants what they hoped to gain from the sessions, emphasizing mutual benefit rather than extractive data collection \cite{Birhane2022, hooks1994, Sloane2022}.

To support multidimensional accessibility, all sessions took place in a community center in the participants’ neighborhood, with food and drinks provided and the seating arranged informally \cite{CostanzaChock2020}. We emphasized that the goal was to share lived experiences and that there were no ‘good’ or ‘bad’ answers. Additionally, the same researcher guided all sessions and ensured that the study payments did not affect the benefits participants received.\footnote{In the Netherlands, there is a maximum amount a resident can earn before it affects government benefits (e.g., sickness benefits).}

Finally, Session 1 centered minoritized standpoints by treating participants’ narratives as the primary source of knowledge about how healthcare and structural inequities are experienced in practice, rather than as one input among many. At the end of the session, participants reflected on the design, suggested changes, and set priorities for subsequent sessions.

\subsubsection{Session 2: Rules From Lived Experience for Participation in Healthcare}

Session 2 extended the focus on lived experience by moving from individual accounts to co-constructed rules, foregrounding the AIM preconditions of a decolonizing stance, co-ownership, methodological flexibility, and centering minoritized standpoints. Session 2 was also inspired by the second phase of BNIM \cite{Wengraf2001}. Like this phase, the session remained grounded in participants’ lived experiences, building directly on the narratives shared in the first session and avoiding external framing. In this sense, it overlapped with BNIM’s narrative follow-up phase, which seeks to further explore and clarify participants’ accounts. However, the second session departed from BNIM in its purpose and structure. Rather than focusing solely on deepening individual narratives, it aimed to foster a collective understanding that could serve as a foundation for subsequent sessions. 

This shift enacted Smith’s (2012) account of decolonizing research, in which communities move from being objects of study to agents of inquiry with authority over what counts as relevant knowledge \cite{smith2012}. It also reflected centering minoritized standpoints: the rules were treated as legitimate normative claims grounded in participants’ positions as women and non-binary people of color in the Dutch healthcare system, rather than as anecdotal input to be later translated by experts \cite{collins2000black, Crenshaw1991}.

Methodological flexibility and co-ownership were built into the session design. Participants could question, rephrase, or reject proposed rules, decide the level of abstraction, from individual encounters to institutional policy, and linger on specific stories when they felt this was necessary before formalizing any rules. The structure was presented as a starting point rather than a fixed protocol, allowing the session to remain responsive to what participants considered most relevant \cite{Udoewa2022}. As in the first session, time was allocated at the end for participants to reflect on the session design and suggest adjustments for subsequent sessions.

\subsubsection{Session 3: If, Where, and How: Lived Experience and Participatory AI}
Session 3 explicitly introduced AI into the process, foregrounding the AIM preconditions of structural availability of refusal, methodological flexibility, and centering minoritized standpoints, while continuing to build on reciprocity and co-ownership. The session began with a brief reflection on the previous two AIM sessions. Lived experience literature defines it as “direct, unmediated experience” that is “local, temporal, embodied, and unique” \cite{Ajjawi2024}. Building on this, we introduced a non-verbal way of expressing lived experience through an exercise inspired by the Rich Picture method \cite{cristancho2019richpictures}. Participants visually represented their lived experiences through drawings informed by prior discussions, emphasizing that there were no right or wrong representations. After completing their drawings, participants could explain them to one another if they wished. 

Following the drawing exercise and building on the reflections and rules identified in earlier sessions, Session 3 structured participants’ agency around three explicit questions: whether AI should play a role at all in responding to the lived experiences shared in earlier sessions and Rich Pictures; if so, where AI should be situated in relation to those lived experiences; and how AI should be designed so that the lived experiences and shared rules from Sessions 1 and 2 remained centered. The session preserved the option for participants to articulate that AI was not an appropriate response, a refusal treated as a legitimate methodological outcome in keeping with the principles of radical participatory design and decolonizing methodologies \cite{Udoewa2022, smith2012, TuckYang2014}. At this stage, we deliberately avoided introducing the AI development life cycle to prevent shaping or constraining participants’ responses; instead, the question was kept open-ended. This sequencing operationalized the argument from Delgado et al. (2023) and Corbett et al. (2023): if participants enter a process whose problem definitions, success criteria, and design artifacts are already set, their contributions cannot reshape what the system is for \cite{Delgado2023, corbett2023power}. Session 3 therefore introduced technical framing only after lived experience had been articulated on its own terms.

Next, we introduced a real-world case of an AI application under development to calculate cardiovascular disease risk \cite{villalobos2025decide} and asked participants how lived experience could be incorporated. After this discussion, the researcher presented Gautam et al.’s (2025) Lived Experience AI Framework (LEAF) as a model that integrates lived experience within the AI development life cycle and addressed participants’ questions \cite{gautam2025experience}. Participants were then asked whether, and if so, where lived experience should be incorporated in such a model. Presenting LEAF here served two purposes: it gave participants a concrete reference for participatory AI and allowed us to compare their responses with our own analytic positioning of lived experience as preparatory rather than integrated. As in previous sessions, this one concluded with time for reflection on both content and process.

\subsubsection{Session 4: Embedding Lived Experience in AI Healthcare Policy}

Session 4 focused on translating participants’ lived experiences, rules, and reflections on AI into the context of public-sector healthcare policy, foregrounding the AIM preconditions of reciprocity, a decolonizing stance, multidimensional accessibility, and co-ownership. In this final session, five policy workers from the municipality where participants live joined both the morning and afternoon sessions (N = 5). Their work focuses on the intersection of healthcare, AI, and participatory approaches involving citizens. The session aimed to co-develop policy guidelines grounded in participants’ lived experiences, aligning with work on participation, accountability, and democratic legitimacy in public-sector AI governance (Zuiderwijk, Chen, and Salem 2021; Wong et al. 2025), while responding to Gilman’s (2023) warning against top-down participation by prioritizing accessible engagement, community expertise, and transparency about how input could affect outcomes.

To maintain a safe and supportive environment, the policy workers joined only after participants and the researcher had reflected on the previous session. This sequencing provided an opportunity to check whether any additional measures were needed to ensure that participants felt comfortable and safe during the interaction and enacted what hooks (1994) describes as honest confrontation: rather than smoothing over the asymmetry of bringing institutional actors into a participant space, the sequencing surfaced it and allowed participants to set the terms on which the encounter would proceed \cite{hooks1994}. We emphasized that participants could choose not to continue if they did not feel comfortable with the conversation. Building on insights from earlier sessions, participants and policy workers then discussed how lived experience could inform AI-related healthcare policy. We emphasized that policy should be adapted to meaningfully integrate participants’ lived experiences, rather than expecting participants to reshape their experiences to fit existing policy frameworks, reflecting a decolonizing stance in which communities are treated as agents of inquiry \cite{smith2012}.

The same community-based location, timing, and food and care provision were maintained to support multidimensional accessibility \cite{CostanzaChock2020}. Co-ownership and reciprocity were reinforced by discussing concrete ways policy workers could apply insights from the sessions, the constraints they faced, and how participants could be involved in future decision-making processes. To close, participants first reflected without policy workers present on how the encounter had felt and what they wanted in future engagements, then joined a joint reflection with policy workers on lessons learned and possible next steps.

\subsection{Experimental Procedure}
Following Design from the Margins \cite{CostanzaChock2020} and intersectionality \cite{Crenshaw1991}, the call for participants intentionally sought women of color and non-binary people of color. This focus is grounded in evidence that these groups have historically experienced minoritization in healthcare, with many services designed by and for men \cite{Obermeyer2019, salles2019genderbias}. In line with an intersectional approach \cite{Crenshaw1991}, our sampling strategy also accounted for race, socioeconomic status, disability, and geographic background.

In total, 13 participants (N = 13) from minoritized communities participated, all of whom rely on the Dutch healthcare system. Participants were recruited through community centers and word of mouth and signed up for either the morning (group 1, N = 8) or afternoon (group 2, N = 5) session series: the two series each comprised four Lived Experience Sessions. The two series ran in parallel, resulting in eight sessions held weekly over four consecutive weeks, each lasting between three and four hours. For Session 4, five policy workers (N = 5) joined the groups: three (N = 3) in the first group and two (N = 2) in the second. Participants remained in the same groups throughout to ensure continuity and preserve the safe space established during the first session \cite{smith2012}.

All sessions were led by the second author, drawing on participatory and feminist research methodologies grounded in researcher positionality \cite{collins2000black, CostanzaChock2020, Crenshaw1991}. While we do not assume that shared identity guarantees trust, this facilitation choice aimed to reduce perceived social distance and power differentials, which prior work shows can inhibit trust and limit the sharing of lived experiences in research with marginalized communities \cite{Sloane2022, CostanzaChock2020}. The first author, who is perceived as a white male, did not lead the sessions; facilitation was led by the second author, a woman of color with lived experience relying on healthcare services. This choice supported a safer research environment and aligned with participatory principles emphasizing reflexivity and relational accountability.

\subsection{Analytical approach}
Our primary aim in this paper is methodological: to introduce AIM as a generalizable stance for the preparatory phase of participatory AI and to illustrate its implementation in one concrete context. We therefore treat the application of AIM as a worked example rather than as an empirical validation of the framework. That is to say, the sessions demonstrate one way of instantiating AIM's preconditions, not a test of whether those preconditions are necessary or sufficient. Our analytical approach is consistent with this aim and with AIM's theoretical commitments for two reasons. First, imposing a predetermined coding scheme on transcripts generated through open-ended, participant-led sessions would reproduce the kind of researcher-imposed framing that AIM is specifically designed to resist \cite{smith2012, collins2000black}. Instead, we read session notes and transcripts iteratively alongside AIM's seven preconditions, identifying moments that were especially illustrative of how those preconditions manifested, or required adaptation, in practice. For instance, when participants redirected the agenda, articulated refusal, negotiated co-ownership, or connected healthcare experiences to broader structural conditions. Second, quotes in the Results section were selected to typify these dynamics and the overall tone of the sessions rather than to surface exceptional cases. They are used to showing how participants expressed their lived experiences in their own terms, and to make visible the interactions through which AIM's preconditions were enacted, tested, and sometimes renegotiated.

\section{Results}

\subsection{AIM Session 1: Context and Narratives}
The first session focused on creating a safe space and building trust among group members and with the session leader. Participants were invited to share their lived experiences in healthcare by means of an open question (SQUIN). All participants shared their lived experiences. For example, one participant explained that she went to the general practitioner (GP), and the first thing he said was: \textit{``So many dark children in Africa have a bulging belly button. I said to my child, `Come on, we're leaving'. I live here in the Netherlands, and then you say things like that''}. Although this lived experience concerned the participant's child rather than her own health, the session leader adapted and followed the participant's narrative flow. Participants shook their heads, saying that they had similar experiences, making clear that, for them, healthcare was always linked to other systems, and lived experiences were always linked to connected ones, never singular: \textit{``My child was very restless and crying very hard. I just went to the GP practice, and I was sitting there, but I wasn't being called because I didn't have an appointment. After a while, the GP came to me and said, `Ma'am, you can't just come here like that''}. After the other participants asked what happened next, the participant explained that she went to another GP: \textit{``I went to another GP, and my child ended up being hospitalized for two weeks with a lung virus… There really are GPs who always want to finish things quickly with me''}. The group then laughed together, saying that it was no coincidence they shared lived experiences. 
One participant described the session as engaging: \textit{``[The session] was surprising and nice to hear everything, well, not nice, but to hear what people go through, to hear their lived experiences''}.

\subsection{AIM Session 2: Rules and Points of Interest}
The goal of this session was to develop rules for healthcare settings guided by patients' lived experiences, shifting the discussion from individual opinions to shared rules and enabling the collective examination of underlying assumptions and practices. As in Session 1, the session design had to remain flexible to secure participant leadership and knowledge creation on their own terms, grounded in their understanding of their local contexts. By centering and following participants' lived experiences, the group co-constructed a set of rules that integrates patients' lived realities into healthcare practice. For example, one participant described the importance of feeling heard and understood, explaining that: \textit{``They do not want to make an effort to truly listen and understand''}. Another participant responded that this goes hand in hand with being taken seriously, to which the group responded: \textit{``Enough with always prescribing painkillers, instead of listening to what we are going through''}. The collaboratively drafted rules generated strong engagement among participants. All participants expressed interest in obtaining copies of the rules they had collaboratively written, emphasizing that: \textit{``[The rules] must be implemented and we [must] submit a petition. That the session should be led toward a good solution''}. In the subsequent sessions, these rules functioned as a participant-authored reference point against which AI-related discussions were tested.

\subsection{AIM Session 3: If, Where, and How: Lived Experience and Participatory AI}

Session 3 aimed to answer if, where, and how AI should be involved in responding to participants' lived experiences. It started by reiterating lived experiences through participant-drawn Rich Pictures, confirming the centralization of lived experience as established through Sessions 1 and 2, and asked participants to determine the role, if any, that AI should play in relation to it. In practice, this session required the most flexibility from the session leader and produced the most unexpected turns of the study. Following the drawing exercise, the session leader explained that participants should take into account their lived experiences, the rules established in Session 2, and their drawings. Participants then engaged in lively debate about whether and how AI should be developed in this healthcare context, while affirming throughout that their lived experiences and the rules they had collectively formulated must shape any AI that emerged from such a process.

This shift proved challenging, as participants described difficulty discussing participatory AI design without concrete knowledge of what such processes entail. As one participant noted, \textit{``It is hard to think about something I do not have knowledge about.''} In response, the session leader introduced a case study of an AI application designed to predict cardiovascular disease and provided an overview of participatory AI, including the life cycle proposed by Gautam et al. (2025). With this background, participants felt more able to engage in discussion, even as they remained hesitant to speak with the same confidence they had shown when discussing healthcare, pointing to the necessity of community-based education and engagement. Interestingly, although the case study provided a concrete example of AI in healthcare, participants frequently grounded their contributions in their own experiences. At the time of the research, a major data breach involving women's cervical cancer screening data had occurred, and all affected individuals received a letter from the government. When participants' concerns about AI centered on this breach, the session leader remained flexible and did not redirect the discussion back to the case study.

Participants engaged in lively debate about centering lived experiences, explaining that their experiences with healthcare and the rules they had collectively formulated should also apply to AI applications in this context. More specifically, participants articulated this positioning in their own terms: \textit{``The problem comes from the lived experiences; the lived experiences are the data and should be the start.''} As one participant noted, \textit{``Lived experiences are local, and systems [like healthcare] are connected to that world.''} This highlights the interconnectedness between healthcare systems, broader social contexts, and the power dynamics that emerge within them. Having made this interconnectedness tangible through sharing their experiences in earlier sessions, participants stressed that understanding both the system (in this case, healthcare) and the experiences it shapes is a necessary step before designing AI intended to improve that system. For example, one participant reflected, \textit{``In a previous session, I explained not feeling completely understood by my healthcare provider; how would AI [in healthcare] do that any better?''} Another added, \textit{``I do not think AI would make me feel comfortable if I do not feel comfortable with my GP in the first place.''} These reflections emerged without prompting from the session leader, connecting healthcare experience directly to AI expectations. Participants therefore argued that \textit{``lived experiences should be investigated thoroughly''} prior to initiating participatory AI design, positioning this work as essential preparatory groundwork. They also emphasized the importance of lived experience in evaluating AI applications: \textit{``We have already encountered all the problems, so the question is: what is the right path to take [for the application], guided by our lived experiences rather than the developers?''} Finally, participants stressed that meaningful participation requires involvement throughout the entire design process: \textit{``You can design all sorts of AI applications, but you first need to do proper research and actively invite people to ask questions and get involved… Make the effort to be inclusive, reach out through mosques, community centers, and other accessible venues, and ask people to collaborate throughout the whole design process.''}

\subsection{AIM Session 4: Embedding Lived Experience in AI Healthcare Policy}
Building on the first three sessions, the fourth session focused on how AI policy in healthcare might incorporate and safeguard the lived experiences of minoritized communities throughout the participatory AI design process. With participants' consent, policy workers working within AI, healthcare, and participatory design were invited to join a structured discussion intended to lay the groundwork for policy guidelines centered on lived experience. However, once the session began, participants quickly shifted the focus toward sharing their own lived experiences with the increasing digitization of healthcare and seeking answers about why the healthcare system, in their view, was increasingly failing those who rely on it. This shift was appreciated by some policy workers, though not all. Nevertheless, the urgency expressed by participants highlighted the importance of direct dialogue between policymakers working on AI and healthcare and minoritized communities. The discussions suggested that policymakers did not always have a concrete understanding of how systemic inequities in healthcare and digitalization were experienced in everyday life. Participants, in turn, described feeling bolstered by the opportunity to engage directly with those they perceived as responsible for shaping these systems: they recorded responses, requested contact information, and sought follow-ups on local initiatives. As one participant put it, \textit{``Come to us, because the people responsible for this do not dare to come here, so come to us more often to talk''}.

Some policy workers expressed frustration with participants' limited technical knowledge of AI. This response underscored a mismatch in expectations: the policymakers appeared to overestimate participants' familiarity with AI technologies while simultaneously underestimating the valuable insights that lived experiences can offer to AI governance, even when expressed in non-technical or non-policy-specific language. Because the session ultimately diverged from its initial structure, no concrete policy guidelines were produced. This divergence was itself a methodological outcome. Our method is not primarily meant to delineate a directly actionable outcome, but to expose the underlying structures (e.g., power imbalances and informational asymmetries) that alienate a socially sound framing of the problem. Nonetheless, both participants and policy workers expressed enthusiasm about the dialogue. 

\section{Discussion}

\subsection{Preconditions are Mutually Constitutive} 
A central observation from the case study is that AIM's preconditions do not operate as independent commitments. Rather, they appeared mutually constitutive across the sessions: each precondition created the conditions under which others could be meaningfully enacted. This has direct implications for how participatory AI frameworks are designed, and specifically for why the preparatory stage AIM proposes cannot be placed within existing participatory AI pipelines. 

The clearest illustration of this interdependence is the relationship between multidimensional accessibility, reciprocity, and the centering of minoritized standpoints. Holding sessions in a community center within participants' neighborhood, providing food that could be taken home, and scheduling at accessible times were not logistical conveniences layered on top of the method. They were the material preconditions that enabled reciprocity. Without them, the relational orientation hooks (1994) describes, a deliberate break from hierarchical research relationships, would have remained aspirational rather than enacted. 
What the sessions make visible, then, is that accessibility did not merely support the other preconditions; it enabled them.

The same holds for the relationship between reciprocity and co-ownership. The opening procedures of Session 1,  discussing participants' rights, asking what they hoped to gain, and addressing concerns before recording began, were not procedural formalities. They were the acts through which co-ownership was established as a relational reality rather than a stated principle. What the sessions illustrate is that shared ownership must be built through repeated relational acts across time \cite{Sloane2022}. When participants in Session 2 requested copies of the co-constructed rules and called for their implementation, this was evidence of that process: participants understood the rules as theirs — authored by them, belonging to them, and actionable by them. That sense of ownership over the output could only emerge if the process that produced it had felt genuinely non-extractive, which is what the relational foundation established in Session 1 made possible. Had the sessions begun within an existing design process, where problem framings and success criteria were already set, there would have been neither the time nor the relational space to build these conditions \cite{Delgado2023, corbett2023power}.

Taken together, these observations 
suggest that interdependence is itself an argument for the preparatory stage: building reciprocity, accessibility, and co-ownership requires sustained relational work that cannot be compressed into a single workshop or retrofitted onto an existing design process. The pre-design stage AIM proposes is therefore not simply one more phase to add to participatory AI; it is the condition under which participatory AI's commitments can begin to be fulfilled, including the commitment that refusal of AI remains a genuine option rather than a rhetorical one.

\subsection{Making Asymmetry a Methodological Outcome}
Session 4 did not produce the policy guidelines it aimed to generate, but we suggest this reading misidentifies its purpose. The session yielded a methodologically significant moment: the asymmetry between institutional and experiential framings of AI became visible to all parties, rather than being smoothed over to serve a predetermined output. This outcome was enabled by the relational foundation established in earlier sessions and by the explicit conditions governing the encounter with policy workers. In their closing reflection without policy workers present, participants linked their willingness to engage, asking questions and requesting contact information, to the sense that the space remained theirs rather than the institution’s. Before policy workers joined, the session leader emphasized that the session existed for participants, not the institution, and that they retained the right to cancel or exit the conversation at any point, exemplifying the structural availability of refusal in the governance encounter itself.

Participants took the opportunity to ask questions of those they saw as responsible for shaping the systems they rely on. As one participant put it: “Come to us, because the people responsible for this do not dare to come here, so come to us more often to talk.” This articulates a directional claim about where the burden of engagement should fall: policy should move toward lived experience, not the other way around. Expecting participants to translate their experiences into technical or policy-specific language reproduces the institutional logic AIM is designed to unsettle and reflects the colonial extraction \citet{Udoewa2022} identifies in participatory practices that retain institutional control over what counts as legitimate knowledge.

The frustration some policy workers expressed with participants’ limited technical knowledge of AI is instructive. It surfaced a previously invisible mismatch in assumptions: policymakers appeared to overestimate participants’ familiarity with AI while underestimating the diagnostic value of lived experience for AI governance, even when expressed in non-technical terms. Rather than resolving this mismatch, the session made it visible to both parties, which is itself a methodological outcome. This points to a limitation of existing participatory AI governance frameworks that evaluate sessions based on the policy outputs they generate \cite{gilman2023democratizing, zuiderwijk2021implications}. AIM proposes that surfacing these asymmetries, and requiring policymakers to move toward lived experience rather than the reverse, is a legitimate and necessary outcome of participatory AI governance, and one that sustained engagement can build upon.

\subsection{From Shared Experience to Collective Agency}
Session 1 created conditions in which people generally shared difficult experiences, recognizing themselves in each other's accounts, finding solidarity in shared hardship. This is in itself not a trivial outcome, but it is not AIM's goal. What the sequencing across sessions was designed to do, and what the empirical material suggests it accomplished, is transforming shared experience into seeds of collective agency. The shift from individual narrative in Session 1 to co-constructed rules in Session 2 is not incidental; it is the methodological move that prevents lived experience from remaining at the level of emotional discharge while ensuring that participants and the session leader work together to form shared intent in a safe, equitable space. This relational foundation does not merely precede the other sessions — it serves as a building block for the knowledge production on which Sessions 2, 3, and 4 depend. Participants did not leave Session 2 having aired their grievances — they left with rules they authored, demanded copies of, and called for implementation. By Session 4, this collective orientation had become explicitly political.

We initially framed Session 4 as producing a dialogical encounter between participants and policymakers. On reflection, however, what the session produced looks less like a dialogue and more like the beginning of accountability politics. Participants did not simply share their experiences with policymakers; they recorded responses, demanded contact information, and sought concrete follow-ups. The rules they had collectively authored in Session 2 started becoming, by Session 4, the terms against which they evaluated what policymakers were doing and not doing. This is not dialogue in any passive sense; it is citizens using collectively produced knowledge as framing to evaluate and hold institutional actors accountable. Whether this accountability orientation can be sustained beyond the sessions is a question for future work, but its emergence within a four-session preparatory stage suggests that AIM's sequencing does more than create a safe space for sharing, a mere dialogical space in which discontent can be expressed. It creates the conditions for participants to position themselves as political actors in relation to the systems that shape their lives.

\subsection{Theoretical Implications}
The order in which knowledge enters the design process determines whose knowledge can shape it. AIM offers an operational response to the critique that participation typically enters the AI design pipeline only after problem framings and design artifacts have already been set \cite{corbett2023power, Delgado2023}. The deeper theoretical claim is not simply that lived experience should be addressed earlier, but that epistemic sequencing, the order in which different forms of knowledge are made available to the design process, is not a logistical question but a political one. Building on \citet{Delgado2023}) demonstration that stakeholders typically enter processes whose terms are already set, and \citet{corbett2023power} identification of problem framing as a site of power, AIM makes visible the mechanism through which this power operates. Prior work on situated interventions for algorithmic equity \citet{Katell2020} and on the conditions under which participation remains meaningful in the age of foundation models \citet{Suresh2024} has laid important groundwork for understanding how context shapes participation; AIM extends this by proposing that epistemic sequencing determines whether minoritized standpoints can reshape what an AI system is for, or only how an existing system is refined \cite{Benjamin2019}.

AIM's preconditions suggest that the normative commitments underlying participatory AI are not independent values to be implemented in parallel, but mutually constitutive conditions that depend on one another to function. The foundation on which AIM rests: participation as justice, decolonizing research, the material conditions of participation, the epistemic authority of minoritized standpoints, and the legitimacy of refusal, has been established across a body of scholarship that this paper draws on throughout \cite{collins2000black,CostanzaChock2020,Sloane2022,smith2012,TuckYang2014}. What AIM's application adds is a relational account of how these commitments interact: reciprocity enables standpoint centering, accessibility enables reciprocity, co-ownership depends on both, and refusal remains meaningful only when the relational infrastructure to support it is already in place. A participatory AI process that implements these commitments as parallel design principles, each treated independently, risks finding that none of them fully functions, including the commitment that refusal of AI remains a genuine option rather than a rhetorical one. The sustained relational work of building these conditions cannot be compressed into a single workshop or distributed across an existing design pipeline \cite{Udoewa2022, Kallina2025}; it requires the preparatory stage AIM proposes.

Session 4 surfaces a principle that existing participatory AI governance frameworks have not yet fully confronted: the burden of bridging institutional and experiential framings of AI should fall on policymakers rather than on participants. This is not a procedural adjustment but an epistemic one, consistent with Collins' (2013) intellectual activism and with the decolonizing stance AIM takes throughout. Work on participatory AI governance has made vital contributions in establishing accountability and democratic legitimacy as central concerns \cite{zuiderwijk2021implications, wong2025key, gilman2023democratizing, gautam2025experience}; what AIM adds is the concrete demonstration that surfacing asymmetry between institutional and experiential framings, rather than resolving it, is itself a governance outcome, and one that sustained preparation of institutional actors, not only of participants, can build upon \cite{Udoewa2022}. AIM can be used by researchers, civil society organizations, institutional actors, and developers working together with minoritized communities. In our case study, inviting policymakers was a deliberate contextual choice for public-sector healthcare AI, not a claim that policymakers are always the right institutional interlocutor. In other settings, for example, inviting AI developers might be more appropriate. However, we emphasize that AIM should be initiated by, or co-initiated with, minoritized communities whose lived experiences are at stake, with institutional actors providing support and resources rather than defining the process unilaterally.

\subsection{Limitations and Future Work}
Despite AIM's goal of making participatory AI design more inclusive by centering lived experiences prior to design processes, we emphasize that no methodological framework can fully eliminate blind spots. Researchers and designers inevitably enter participatory processes with their own assumptions, institutional constraints, and implicit goals, which shape how participation is structured and interpreted. While centering lived experience can help surface perspectives that are often minoritized in AI development, it does not guarantee completeness or representativeness, nor does it remove existing power asymmetries. We therefore stress that AIM should not be understood as a definitive solution for participatory AI, but rather as a reflexive preparatory framework. Ongoing critical reflection, by researchers, designers, policymakers, and participants, remains essential to identify whose voices may still be missing, how participation is being framed, and how emergent insights are translated into AI design and governance. We recognize that this approach may not be feasible for all AI applications but may be especially relevant for applications developed for wide public use, such as government healthcare AI applications.

Although AIM is grounded in the lived experiences of minoritized communities and adopts a bottom-up approach, it was developed and evaluated within a healthcare context. As such, a limitation of this work is that the method's applicability to other domains remains unexamined. Future work should therefore investigate how AIM translates to different sociotechnical settings. Additionally, because the method centers lived experience,  it is inherently collaborative and iterative; this paper  presents an initial instantiation rather than a finalized  approach, and our engagement with participants continues  beyond the study reported here. We have committed to sharing results in accessible formats, inviting participants’ feedback on the evolving method, and exploring further collaborations around local healthcare and AI initiatives.

\section{Conclusion}
This paper introduced AI From the Margins (AIM) as a methodological stance for the preparatory phase of participatory AI design in healthcare. Rather than prescribing a fixed protocol, AIM articulates preconditions that can guide the redesign of participatory AI methods. By positioning lived experience as a preparatory foundation rather than a supplementary input, AIM connects theoretical commitments to inclusive and decolonizing AI with the practical conditions required for meaningful participation. Findings from a healthcare case study, involving eight Lived Experience Sessions with 13 women and non-binary people of color, illustrate how AIM’s preconditions can be enacted through specific techniques. This instantiation also shows that the preconditions, rather than the techniques themselves, are what travel: epistemic sequencing is not a logistical detail but the site at which the politics of participatory AI are decided.
\bibliography{references}

\end{document}